\documentclass[journal=jpcbfk,manuscript=article]{achemso}
\usepackage{graphicx}
\usepackage{amsmath}
\usepackage{amssymb} 
\usepackage[usenames, dvipsnames]{color}
\newcommand{\cF}{\mathcal{F}}
\newcommand{\onlinecite}[1]{\hspace{-1 ex} \nocite{#1}\citenum{#1}} 
\DeclareMathOperator\erfc{erfc}

\author{Matheus Girotto}
\email{matheus.girotto@ufrgs.br}
\affiliation{Instituto de F\'isica, Universidade Federal do Rio Grande do Sul, Caixa Postal 15051, CEP 91501-970, Porto Alegre, RS, Brazil}

\author{Thiago Colla}
\email{colla@iceb.ufop.br}
\affiliation{Instituto de F\'isica, Universidade Federal de Ouro Preto, CEP 35400-000, Ouro Preto, MG, Brazil}

\author{Alexandre P. dos Santos}
\email{alexandre.pereira@ufrgs.br}
\affiliation{Instituto de F\'isica, Universidade Federal do Rio Grande do Sul, Caixa Postal 15051, CEP 91501-970, Porto Alegre, RS, Brazil}

\author{Yan Levin}
\email{levin@if.ufrgs.br}
\affiliation{Instituto de F\'isica, Universidade Federal do Rio Grande do Sul, Caixa Postal 15051, CEP 91501-970, Porto Alegre, RS, Brazil}

\title{Lattice model of an ionic liquid at an electrified interface}
\begin{document}

\begin{abstract}

We study ionic liquids interacting with electrified interfaces. 
The ionic fluid is modeled as a Coulomb lattice gas.  
We compare the ionic density profiles calculated using a
popular modified Poisson-Boltzmann equation with the explicit Monte Carlo simulations. 
The modified Poisson-Boltzmann theory fails to capture the structural features of the double layer and is also unable to correctly predict the ionic density at the electrified interface. 
The lattice Monte Carlo simulations qualitatively capture the coarse grained structure of the double layer in the continuum.
We propose a convolution relation 
that semiquantitatively relates the ionic density profiles of a continuum ionic liquid 
and its lattice counterpart near an  electrified interface. 


\end{abstract}

\maketitle

\newpage

\section{Introduction}

Room temperature ionic liquids (RTILs) have attracted a lot of attention over the last two decades. The urgent need for efficient energy storage makes RTILs appropriate technologically for renewable energy devices\cite{Wi09,LiLi09,HeYa14}. RTILs present high capacitance and fast charging\cite{LiWu15} when interacting with electrified interface (EI), substituting dielectric materials in supercapacitors and ultracapacitors\cite{SiGo08,SiGo10,JiWu16,LiFe17}. Unlike traditional solvent-based electrolytes, RTILs are stable under a larger voltage window, making them ideal for applications involving storage devices and electrolyte-gated supercapacitors.  Furthermore, understanding the behavior of RTILs at EIs can help in the development of electroactuators\cite{CoZh10,WiZh13}, which could be used as artificial muscles.
Theoretically, RTILs stand at the frontier of statistical mechanics\cite{Fr11}. They can be seen as either solvent-free electrolytes or dense ionic plasmas.  RTILs exhibit strong inter-ionic interactions, providing a practical example of a strongly correlated Coulomb system, interesting on purely fundamental grounds \cite{Le02,ScWe04,HeWu14}. 

Theoretical analysis of electrochemical systems makes use of the concept of Electric Double Layers (EDLs), a notion first introduced by Helmholtz when studying electrolytes interacting with EIs\cite{He53}. EDL consists of EI and a neutralizing layer of counterions and coions\cite{HeWu14,LiWu16}. The concept has been used with success for understanding and developing electrical devices\cite{JiWu13,JiHe11}. The canonical approach to understand EDLs is that of Gouy and Chapman (GC)\cite{Pa90}. GC theory predicts an exponential decay of the potential from the EI and a unique differential capacitance minimum for the applied voltage. However, this is not what has been observed experimentally for RTILs\cite{FeKo14}. This failure is due to the mean field character of GC theory, which is valid only for dilute room temperature aqueous electrolytes. RTILs at EIs can present several layers of oscillatory charge density\cite{ChRo14}, making the classical concept of uniform EDL somewhat misleading. A number of sophisticated approaches have been tried to overcome the shortcomings of GC theory, including Density Functional Theory (DFT)\cite{OuBh83,WuLi07,PaGh94,Gr87,HeWu14} which attempts to take into account both size and the electrostatic correlations between the ions of the system \cite{Le02}.  The DFT theory, however is numerically demanding and very difficult to extend to strong coupling regime relevant for RTILs.  One relatively straightforward approach which
has been extensively used in the RTIL community\cite{FeKo14}  relies on a modified Poisson-Boltzmann (mPB) equation, which approximately accounts for the finite size of both ions and solvent.  

\begin{figure}[t]
\begin{center}
\includegraphics[scale=0.4]{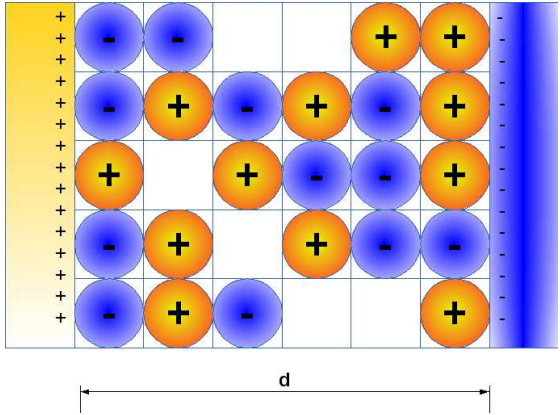}
\end{center}
\caption{Lattice model of RTIL at an EI. The compacity parameter for this $2$D example is $\gamma=\frac{5}{6}$.}
\label{fig1}
\end{figure}

An accurate description of RTILs is a very challenging task. This is mainly due to the very high packing fraction and strong electrostatic interactions inherent to these systems, resulting in a failure of traditional liquid-state based theories. Furthermore, a large variety of experimental synthesis techniques makes it very difficult to model these systems in a unified way \cite{Hay15}. Despite these  difficulties, a lot of effort has been recently invested in an attempt to theoretically characterize the most relevant aspects of these complex systems \cite{Kon11,Hay15,Ila16,LiWu16,Lian16,Liu16,Lian16_2,Kong15,Ila16,Var16,Dud16,Cheng17}. Absence of an aqueous solvent results in  an enhancement of the electrostatic coupling, so that positional correlations play a very important role in determining the structure of the EDLs. As a consequence, the ionic profiles of  RTILs  in the vicinity of charged surfaces feature a layering-like behavior which strongly contrasts with the monotonic diffuse distributions predicted by the usual mean-field approaches \cite{JiHe11}. A feasible  alternative to continuum theories is to use lattice models to study highly concentrated systems. Although very simplified in nature, lattice models can provide valuable physical insights into the underlying mechanisms of many complex systems. Lattice models are also very appealing for Monte Carlo (MC) simulations, since when performing continuum simulations acceptance rate of large moves is usually very small due to hard core repulsion between the ions.  A great deal of effort has been put in extending the traditional liquid-state theories to concentrated lattice systems \cite{Her02,Laf03_2,Laf03,Laf04,Sid09,Bak12,WuBhu17}. An example is the aforementioned mPB  theory, also referred to as Fermi-Poisson or lattice-Poisson-Boltzmann approach \cite{FeKo14,FeKo17}. It is the natural extension of the traditional Poisson-Boltzmann (PB) theory to lattice gases, whereby each ion is allowed to occupy one single lattice site whose typical dimension scales with the ionic size. This approach has an advantage of partially taking into account the finite-size effects in a very simple and intuitive way, since particles are not allowed to occupy the same lattice site \cite{Kor07}. In the case of electrolyte solutions, it is also important to account for the solvent-ion exclusion effects, since the overall packing fraction is kept constant. The question of how accurate this mean-field mPB is in comparison with its continuum PB counterpart seems to be strongly dependent on the particular set of system parameters. In the case of highly charged surfaces, it corrects the simple U-shaped behavior of the differential capacitance curve predicted by the PB theory \cite{FeKo14,Kor07,FeKo08,FeKo082}. On the other hand, at very high salt concentrations it predicts an incorrect saturation of the counterion profile near a charged surface, which obviously contradicts the contact-value theorem \cite{Fry12}.
One can argue, however, that since the mPB equation was constructed on the basis of a lattice gas free
energy, it might not provide a good description of a continuum ionic liquid, but could provide a reasonably
accurate description of its lattice counterpart.  Therefore in order to explore the limitations of this equation,
it is important to compare it directly to the exact model on which it is based, {\it i.e.}, Coulomb lattice gas in contact with EI, see Fig.~\ref{fig1}. 

The purpose of this paper is twofold.  First we analyze the mPB equation by comparing its predictions directly with the  MC solution of the {\it same} lattice model on which it is based.  We then 
explore if a lattice model can capture important features of a continuum ionic liquid.  

\section{The modified Poisson-Boltzmann theory}

We now briefly review some of the main aspects of the PB theory extended to lattice gases. For further details, we refer the reader to more specific literature on this topic \cite{Grim47,Bor97,Kor07,Tre08,Cap03,FeKo14}. We consider the lattice system depicted in Fig.~\ref{fig1}, whereby each lattice site can be either empty or occupied by one single ionic specie $i=\pm$. Each individual site can be viewed as a coarse-grained charge distribution of the corresponding continuum system, averaged over a region whose size is proportional to the particle core. In the case of solvent-based electrolytes, empty sites represent the solvent molecules, whereas in the case of RTILs they stand for the voids present within the liquid. Let $\eta_{i}(z)$ be the probability that a given lattice site whose center is at a lateral distance $z$ (in terms of the corresponding lattice site spacing $R$) from the charged plate is occupied by an ion of component $i$. For continuum systems this quantity can be clearly identified with the local volume fraction of component $i$ at this position. Since all lattice sites are necessarily occupied or empty, the following normalization condition must hold:
\begin{equation}
\sum_{i}\eta_{i}(z)+\eta_{s}(z)=1,
\label{eqt1}
\end{equation}    
where $\eta_{s}(z)$ represents the probability of the site being unoccupied. For solvent-based electrolytes, this quantity can be related with the local solvent volume fraction. In the mean-field approach, the Helmholtz free energy per unit area $\cF$ can be split into an ideal and mean electrostatic contributions. Assuming the limit in which the system size is much larger than the typical site dimension $R$, the sum over the sites can be replaced by integrals over continuous grids. In this limit, the entropic contribution to the free energy per unit of transverse area can be written as:
\begin{equation}
\beta \cF^{\mathrm{id}}=\sum_{i}\int_{0}^{\bar d}\eta_{i}(z)\left[\ln(\eta_{i}(z))-1\right]dz+\int_{0}^{\bar d}\eta_{s}(z)\left[\ln(\eta_{s}(z))-1\right]dz,
\label{eqt2}
\end{equation}  
where $\beta=1/k_B T$ is the inverse thermal energy and $\bar d=d/R$ is the dimensionless distance between the plates (see Fig.~\ref{fig1}). The interaction energy (per unit of transverse area) within the mean-field approach is:
\begin{equation}
\beta \cF^{\mathrm{el}}=\beta q\sum_{i}\alpha_i\int_{0}^{\bar d}\eta_{i}(z)\phi_p(z)dz+\dfrac{\beta q^2}{2\epsilon}\sum_{ij}\int{\dfrac{\alpha_i\eta_i(z)\alpha_j\eta_j(z')}{|\mathbf{r}-\mathbf{r}'|}dz d\mathbf{r}'},
\label{eqt3}
\end{equation}  
where $q$ is the charge of a proton, the $\alpha_i$'s represent the valencies of ionic species, $\epsilon$ is the dielectric constant and $\phi_p(z)$ is the electrostatic potential produced by the EIs. The first term on the right-hand side stands for the ion-wall interaction, whereas the second term represents the mean-field electrostatic interaction among the ionic components (the second integral being performed over all lattice sites).\footnote{Strictly speaking, the total electrostatic energy should also comprise an electrostatic wall-wall contribution, which is omitted here since it has no influence on the resulting Euler-Lagrange relations.}

In equilibrium the occupation probabilities $\eta_{i}(z)$ must minimize the total free energy  per unit of area $ \cF =\cF^{\mathrm{id}}+\cF^{\mathrm{el}}$ under the constraint (\ref{eqt1}). Using the above mean-field functional, it is easily found that the  occupation distributions satisfy the following relation:
\begin{equation}
\eta_{i}(z)=\left[1-\sum_{j}\eta_j(z)\right]\tilde{\eta}_{i}e^{-\beta q\alpha_{i}\phi(z)},
\label{eqt5}
\end{equation}
with $\phi(z)$ representing the total electrostatic potential at a lateral position $z$ and $\tilde{\eta}_{i}\equiv e^{\mu_{i}}$, with $\mu_i$ being the Lagrange multipliers used to enforce 
particle conservation within the system.  In the present planar geometry, this quantity reduces to
\begin{equation}
\phi(z)=-\dfrac{4\pi \sigma_s z}{\epsilon}-\dfrac{4\pi q}{\epsilon}\sum_i \alpha_{i}\int_{0}^{z}\eta_{i}(z')(z-z')dz',
\label{eqt4}
\end{equation} 
where $\sigma_s$ denotes the surface charge density on the electrodes. The first term on the right hand side of the above relation is the electrostatic potential provided by the charged electrodes, whereas the second term stands for the mean-field potential resulting from the inhomogeneous ionic distributions. Here we set the potential to vanish at contact with the first plate, $\phi(z=0)=0$.

If we now define the overall occupation probability $\eta_{o}(z)$ of the site at position $z$ as:
\begin{equation}
\eta_{o}(z)=\sum_{i}\eta_{i}(z),
\label{eqt6}
\end{equation}
it follows directly from Eq.~(\ref{eqt5}) that this quantity can be written as:
\begin{equation}
\eta_{o}(z)=\dfrac{\sum_{i}\tilde{\eta}_{i}e^{-\alpha_{i}\Phi(z)}}{1+\sum_{j}\tilde{\eta}_{j}e^{-\alpha_j \Phi(z)}},
\label{eqt7}
\end{equation}
where $\Phi(z)\equiv \beta q \phi(z)$ is the dimensionless electrostatic potential.
In the case of RTILs, this relation describes the probability of finding the lattice site at $z$ occupied by {\it any} ionic component. Since this relation should hold for every lattice site $z$, it follows from the definition (\ref{eqt6}) that the individual occupation probabilities $\eta_{i}(z)$ should satisfy:
\begin{equation}
\eta_{i}(z)=\dfrac{\tilde{\eta}_{i}e^{-\alpha_{i}\Phi(z)}}{1+\sum_{j}\tilde{\eta}_{j}e^{-\alpha_j \Phi(z)}}.
\label{eqt8}
\end{equation}
Notice that the structure of the ionic occupation distributions resembles the energy level distribution of an ideal Fermi gas. In the limit where all the lattice sites become occupied ($\eta_{s}(z)=0$ in Eq.~(\ref{eqt1})), the continuum PB distribution is recovered. The coefficients $\tilde{\eta}_i$ can be determined from the overall number of ions within the system. 

Let us now consider the special case of a symmetric binary ionic mixture. In this situation, the occupation probability of cations and anions should satisfy the symmetry condition $\eta_{+}(z)=\eta_{-}(\bar d-z)$ (remember that the plates are oppositely charged, see Fig.~\ref{fig1}). In particular, at the mid-plane the condition $\eta_{+}(\bar d/2)=\eta_{-}(\bar d/2)\equiv \eta_M$ should be fulfilled.  In terms of this single parameter, the occupation distributions for a binary electrolyte take the form:
\begin{equation}
\eta_{\pm}(z)=\dfrac{\eta_{M}e^{\mp\delta\Phi(z)}}{1+2\eta_{M}\left[\cosh(\delta\Phi(z))-1\right]},
\label{eqt10}
\end{equation}
where $\delta\Phi\equiv\Phi(z)-\Phi_M$, with $\Phi_M$ being the mid-plane electrostatic potential. 
The above distributions can be solved numerically together with Eq.~(\ref{eqt4}) with a simple iterative algorithm. In practice, it is the compacity parameter $\gamma=(N_{+}+N_{-})/N$ (with $N$ being the total number of lattice sites), rather than the middle-plane occupations $\mu_M$, which is fixed in our MC simulations. We therefore have to adjust the parameter $\eta_M$ until the relation
\begin{equation}
\gamma=\sum_{i}\int_{0}^{\bar d}\eta_{i}(z)dz
\label{eqt11}
\end{equation}
for the desired compacity parameter is fulfilled. \\

\begin{figure}[ht]
\centering
\includegraphics[width=8cm,height=5.5cm]{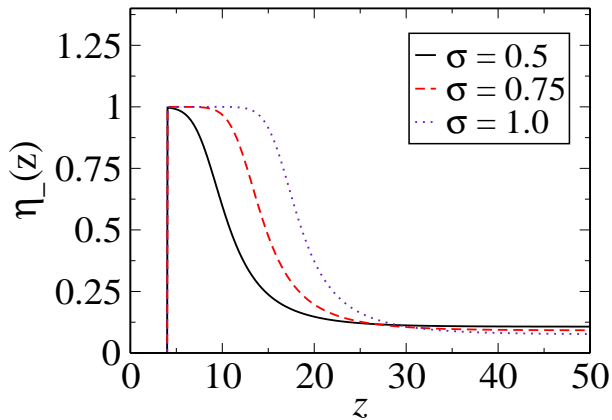}
\caption{Density distributions of anion close to a positively charged surface. The compacity parameter is $\gamma=0.25$, and the lattice spacing is $R=8$\AA. The dimensionless surface charge density is defined as $\sigma=\sigma_s R^2/q$. As the surface charge increases, the counterion contact density does not grow indefinitely, as is required by contact theorem for continuum ionic liquid, but saturates as all the lattice layers close to the surface become occupied by the counterions.}
\label{sat}
\end{figure}

A closer look at Eq. (\ref{eqt10}) reveals a very peculiar property of the mPB occupancy distributions. Lets consider for instance the anionic distribution $\eta_{-}(z)$ close to a (highly) charged positive  plate, see Fig.~\ref{sat}. In this limit we  must have $\beta\delta\Phi=\beta(\Phi(z)-\Phi_M)\gg1$, in such a way that $\eta_{-}(z)$ should approach unity. We therefore conclude that {\it all} lattice sites in the vicinity of the charged plate (where $\beta\delta\Phi\gg1$) will be occupied by neutralizing counterions. This means that the density distribution of counterions in this regime should saturate close to the charged plates, reaching a steady regime of constant charge density, as can be seen in Fig.~\ref{sat}. As the first layer of sites close to the surface becomes fully occupied, the counterions begin to fill in the neighboring layers due to their strong attraction to the charged wall. This contrasts with traditional non-modified PB equation, for which the contact density grows indefinitely as the surface charge density becomes increasingly large.  The mPB equation also seems to contradict the well-known result from the contact-value theorem, which states that for {\it continuum systems} counterion concentration at contact with a charged plate should scale with the square of the surface charge \cite{Hen78,Hen79,Des01}. In spite of this apparently unrealistic behavior when applied to {\it continuum systems}, the mPB equation seem to correctly reproduce the qualitative behavior of the electrolyte differential capacitance in the regime of highly charged capacitors, improving over GC results~\cite{FeKo14} . This apparent discrepancy demands a deeper analysis on how accurate the mPB predictions really are. In the next sections, we aim to perform such analysis by comparing mPB with MC lattice and continuum results.   
\vspace{0.25cm} 

\section{Model and Monte Carlo Simulations}

Computer simulations of ionic liquids pose a number of challenges \cite{Mer13}. It is essential to properly account for the long-range nature of the Coulomb potential -- which is computationally very expensive\cite{BoHe00,ToBo96}. Also, due to charge density oscillations near the EI, the simulation box must be made sufficiently large to achieve the bulk-like regime far from the EI \cite{FeKo14}. Finally, the 2D characteristic geometry of the problem slows down even further the simulations, requiring more specialized techniques necessary to study slab geometry \cite{YeBe99,KaMi01,ArDe02,Pe79,Ad97,Cl97,Sp97}. An alternative is to employ coarse-grained models to partially overcome some of these drawbacks \cite{Brei14_1,Brei14_2}. One such approach is a lattice gas model in which ions are constrained to occupy discrete positions inside the simulation cell.  Here we will use a recently developed Monte Carlo algorithm \cite{GiLe16}, which treats the field produced by the EI on the ions as an external potential, saving considerable amount of computational time. We will perform both MC simulations of a continuum ionic liquid and of its lattice counterpart\cite{KoFi02,RaSh93,SeCo05,DaBi15}.  Unfortunately, due to low move acceptance rates, continuum MC simulations can only be performed for sufficiently low ionic couplings and volume fractions, which are not realistic for ionic liquids. On the other hand, the Coulomb lattice gas model can be relaxed to equilibrium even for  parameters appropriate for real ionic liquids. In spite of this difference, 
the comparison between continuum and lattice systems
should provide us with valuable insights regarding the applicability of discrete models to study real ionic liquids.

Our simulation cell has volume $V=L_xL_yL_z$, with $L_x=L_y=L=80$\AA\ and $L_z=6L$. The ionic liquid is confined to 
the region $-L/2 < x < L/2$, $-L/2 < y < L/2$ and $0 < z < L_z/2$. The positively charged EI is positioned at $z=0$ and the negative EI is at $z=d$, with $d=L_z/2=240$\AA. In the region $-L_z/2 < z < 0$ there is vacuum.
The Bjerrum length, defined as $\lambda_B=q^2/\epsilon k_B T$, will assume in the simulations two values, $\lambda_B=7.2$\AA\ and $\lambda_B=38.4$\AA. The former value is appropriate for electrolytes dissolved in pure water at room temperature. The latter $\lambda_B$ is suitable for RTILs with dielectric constant $\epsilon=\epsilon_{IL}=15$, which is a typical value for a large number of RTILs\cite{WaWe03,Ko08,DrWe11}. The particles are considered as hard spheres with ionic radii $a=4$\AA\ with charges $q$ for cations and $-q$ for anions. The EIs carry surface charge density of $\sigma_s=\mp 0.04$C/m$^2$ or $\sigma_s=\mp 0.10$C/m$^2$.

The total electrostatic potential energy of a charged system containing $N$ ions of charges $q_j$ and two infinite charged plates was calculated in Refs.~[\onlinecite{DoLe14}] and  [\onlinecite{GiLe16}]. To properly 
account for the long-range nature of the Coulomb force one can not use simple minimum image periodic boundary conditions.  Instead the system is periodically replicated and the energy is calculated by summing over all the replicas. To efficiently perform this summation, the electrostatic potential is split into long and short-range contributions - plus a reminiscent self-interaction potential. 
The short range electrostatic potential is
\begin{equation}\label{phi_short}
\phi_S({\pmb r})=\sum_{j=1}^{N} q_j\frac{\text{erfc}{(\kappa_e |{\pmb
r}-{\pmb r}_j|)}}{\epsilon |{\pmb r}-{\pmb r}_j|} \ ,
\end{equation}
where ${\pmb r}_j$ is the position of charge $q_j$ and ${\pmb r}$ is where the potential is calculated. The damping parameter $\kappa_e$ is set to $\kappa_e=5/L$.
The long-range electrostatic potential can be written as~\cite{DoLe14,GiLe16}
\begin{eqnarray}\label{phi_long2}
\phi_L({\pmb r}) = \sum_{{\pmb k}\neq{\pmb 0}}\frac{4\pi}{\epsilon V |{\pmb k}|^2}
\text{exp}\left(-\frac{|{\pmb k}|^2}{4\kappa_e^2}\right)\sum_{j=1}^{N}q_j\text{exp}[i{\pmb k}\cdot({\pmb r}-{\pmb r}_j)] - \frac{2\pi}{\epsilon V}\sum_{j=1}^Nq_j(z-z_j)^2 \ .
\end{eqnarray}
The number of ${\pmb k}$-vectors defined as ${\pmb k}=(2\pi n_x/L,2\pi n_y/L,2\pi n_z/L_z)$, where $n's$ are integers, is set to around $10^4$ in order to achieve convergence.
The potential produced by the EIs is
\begin{equation}
\phi_p({\pmb r})= -\frac{4\pi}{\epsilon}\sigma_s z \ ,
\end{equation}
and the self-interaction potential is
\begin{equation}
\phi^{\mathrm{self}}({\pmb r_i})= q_i \lim_{{\pmb
r} \rightarrow {\pmb r}_i}\frac{\text{erf}{(\kappa_e |{\pmb
r}-{\pmb r}_i|)}}{\epsilon |{\pmb r}-{\pmb r}_i|} \ .
\end{equation}
The total energy assumes the form~\cite{GiLe16}
 \begin{equation}\label{Ue}
 \begin{split}
 U=\dfrac{1}{2}\sum_{i,j}^N\dfrac{q_iq_j\erfc(\kappa\textbf{r}_{ij})}{\epsilon\textbf{r}_{ij}} + \dfrac{2\pi}{V}\sum_{\textbf{k}\neq0}\dfrac{|S(\textbf{k})|^2}{\epsilon k^2}e^{-k^2/4\kappa^2} \\
  - \dfrac{\kappa}{\epsilon\sqrt{\pi}}\sum_i^Nq_i^2 + \dfrac{2\pi}{\epsilon V}(M_z^2 - \sum_i^Nq_i\sum_j^Nq_jz_j^2) + \sum_i^N q_i\frac{4\pi}{\epsilon}\sigma_s z_i \,
 \end{split}
 \end{equation}
 where $S(\textbf{k})=\sum_j^Nq_je^{i\textbf{k}\cdot\textbf{r}}$ and $M_z$ is the total electric dipole moment in the $z$ direction. The second term between parenthesis is zero for a system containing oppositely charged surfaces - which is the case in the present paper.
 
To perform MC simulations we used Metropolis algorithm with $10^4$ MC steps to achieve equilibrium. The profile averages were performed with $3\times10^4$ uncorrelated samples and were made counting the average number of particles in bins of volume $\Delta zL^2$ along the $z$ direction. The value of $\Delta z$ was set to $0.5$\AA\ in continuum model and $R$ in the lattice model. In order to sample appropriately the phase space, we performed swaps between the cations and anions and between the particles and the vacancies in the lattice model. We note that this technique obeys detailed balance\cite{KoFi02}. For the lattice Coulomb gas, the lattice spacing corresponds to particle  diameter.

\section{Results and Discussions}

\begin{figure}[ht]
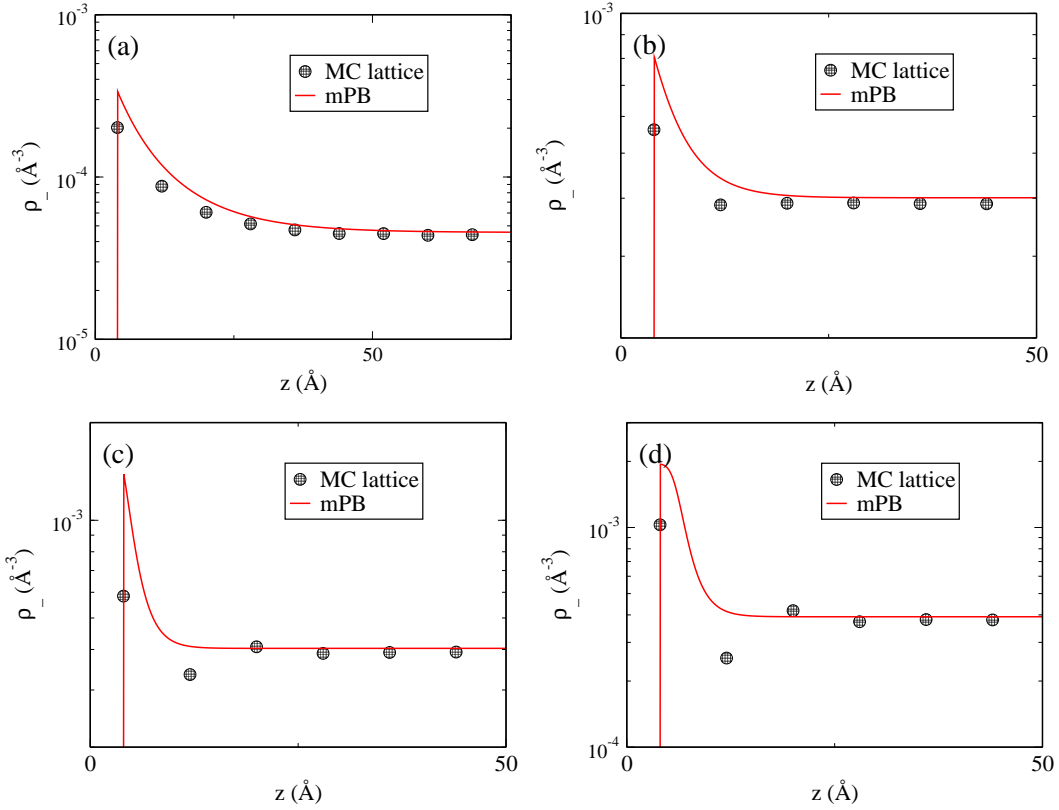

\begin{center}
\includegraphics[scale=0.25]{fig2a.eps}\hspace{0.3cm}\vspace{0.3cm}
\includegraphics[scale=0.25]{fig2b.eps}\hspace{0.3cm}
\includegraphics[scale=0.25]{fig2c.eps}\hspace{0.3cm}
\includegraphics[scale=0.25]{fig2d.eps}\hspace{0.3cm}
\caption{Density profile of anions near the positively charged EI. The continuous line is the result of mPB while symbols are MC simulation of a Coulomb lattice gas. The surface charge at the electrode is $\sigma_s=0.04$C/m$^2$  in (a), (b) and (c) and $\sigma_s=0.10$C/m$^2$ in (d). The lattice spacing is $R=8$\AA.  The parameters are: (a) $\lambda_B=7.2$\AA~ and $\gamma=\frac{1}{20}$. (b) $\lambda_B=7.2$\AA~ and $\gamma=\frac{4}{10}$. (c) and (d) $\lambda_B=38.4$\AA~ and $\gamma=\frac{4}{10}$.}
\label{fig2}
\end{center}
\end{figure}

We begin by comparing the density profiles calculated using MC simulation of the lattice Coulomb gas with the predictions of mPB equation.  The counterion density profiles on  {\it logarithmic scale} are shown in Figs.~\ref{fig2}. For weak couplings (small Bjerrum length) and  low ionic concentrations the mPB equation describes quite accurately the simulation results, see Fig.~\ref{fig2}(a).  We note, however, that for such parameters the usual (unmodified) PB equation is equally accurate. As we increase the compacity factor, while keeping the Bjerrum length $\lambda_B$ constant, the mPB theory begins to breakdown, see Fig.~\ref{fig2}(b). This clearly indicates that the theory is incapable to properly capture the finite-size effects that emerge at large concentrations, and result in a non-monotonic structure of the EDL. When the simulations are performed for parameters typical of RTILs, see Fig.~\ref{fig2}(c), the theory performs very poorly, being unable to account for the complex EDL structure,
predicting the contact density off by an order of magnitude. For moderately high surface charges, see Fig.~\ref{fig2}(d), the theory presents a saturation that is not seen in simulations and does not capture the layering structure near the EI. This failure is due to the mean-field character of mPB equation which does not take into account electrostatic correlations between the ions and treats steric repulsion at a local density level, which is known to be insufficient for dense fluids. The adaptation of more precise theories\cite{Her02,Laf03_2,Laf03,Laf04,Sid09,Bak12,JiHe11,MeWu11,JiWu11} based on non-local weighted density approximations is, therefore, necessary in order to properly account for the thermodynamics of
a Coulomb lattice gas at high density and in the strong coupling limit appropriate for modeling RTILs.  
\begin{figure}[h]
\begin{center}
\includegraphics[scale=0.25]{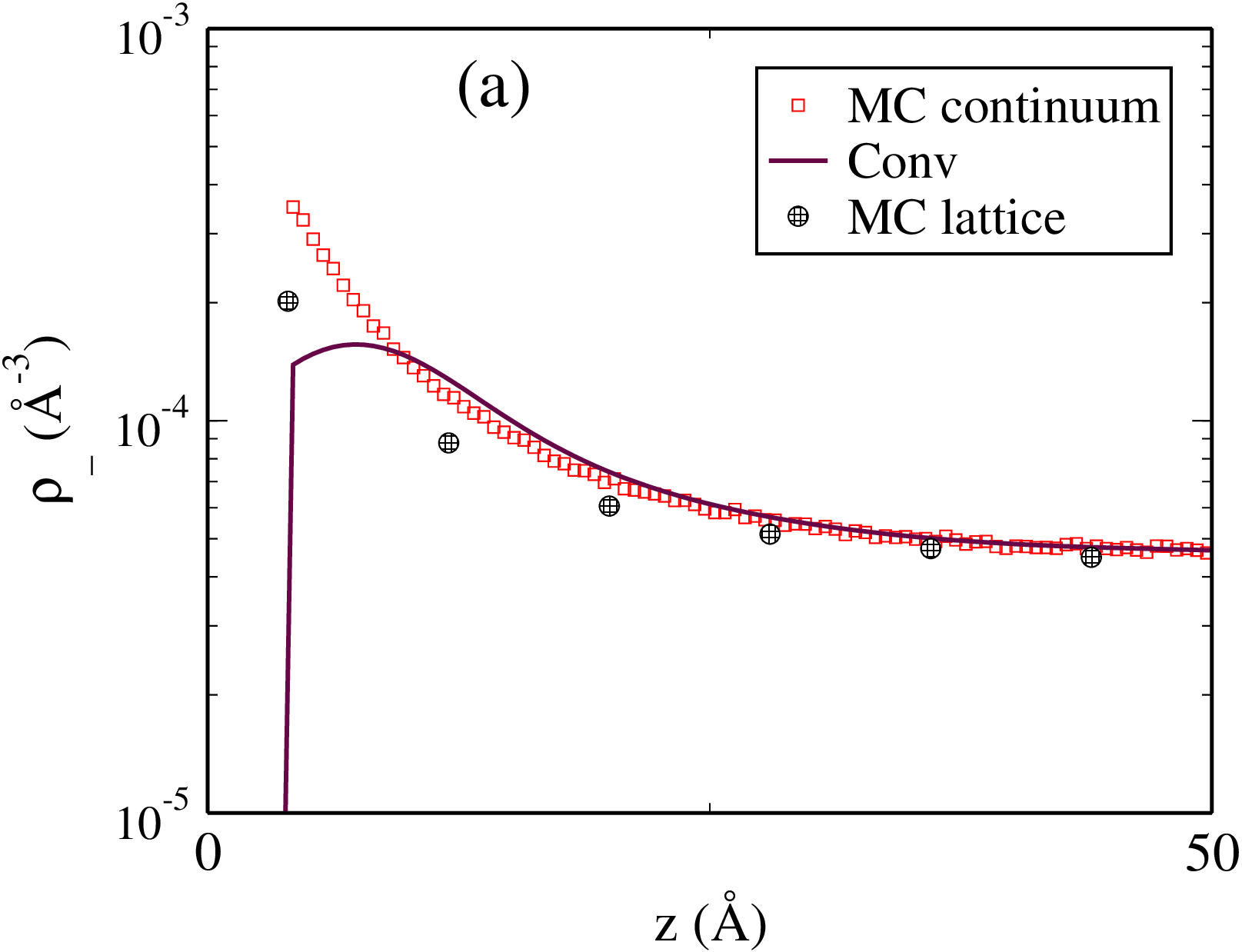}\hspace{0.3cm}
\includegraphics[scale=0.25]{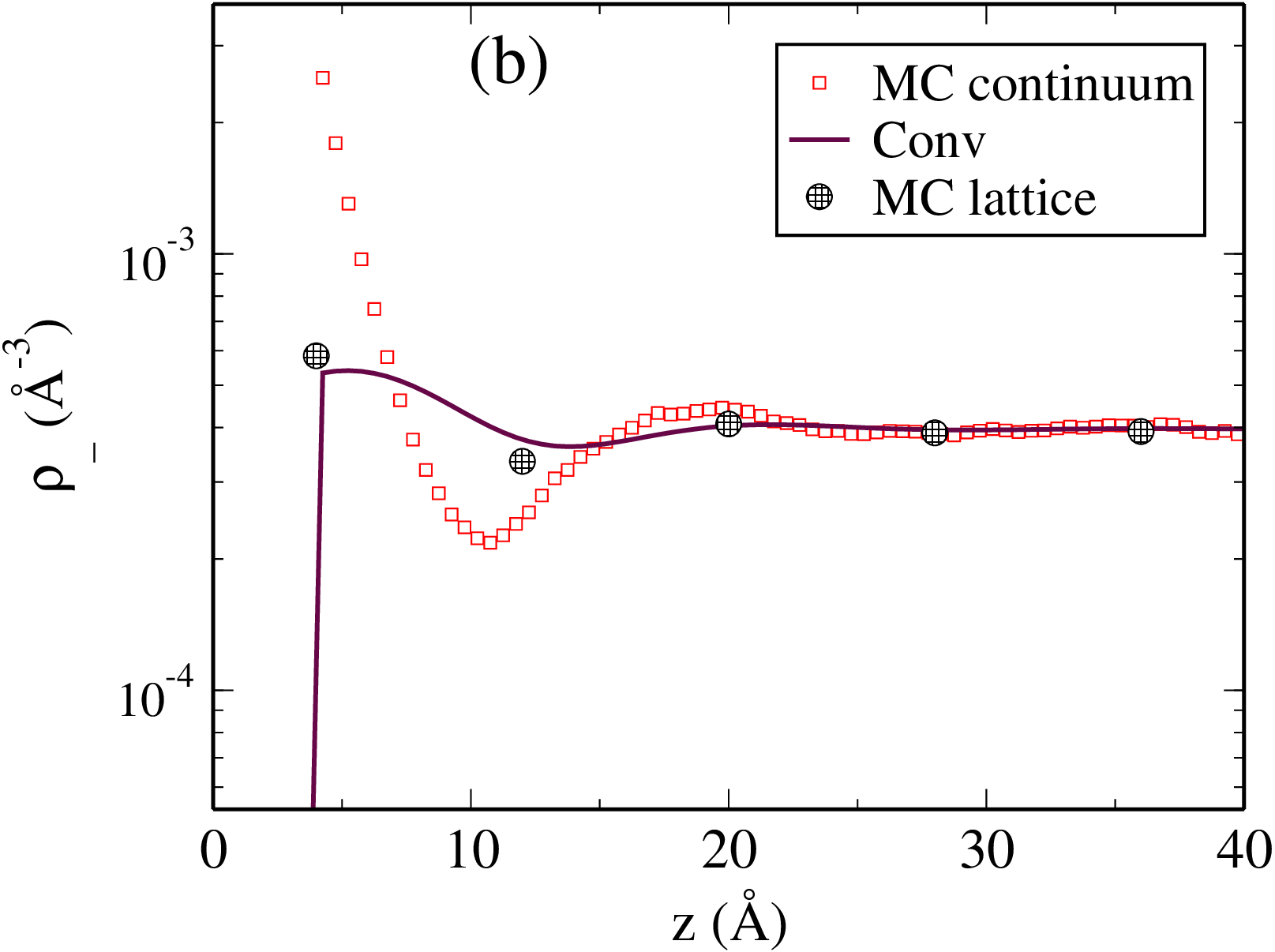}
\end{center}
\caption{Density profiles of anions near a positively charged EI. Square symbols are MC simulations of a continuum ionic fluid, while circles are the results of simulations of the Coulomb lattice gas. The line corresponds to the convolution of continuum MC profile with a Gaussian, see Eq.~(\ref{cov}). The surface charge on the electrode is $\sigma_s=0.04$C/m$^2$ . The parameters were chosen to correspond to (a) dilute electrolyte, $\lambda_B=7.2$\AA~ and $\gamma=\frac{1}{20}$, and (b) to typical RTILs, $\lambda_B=38.4$\AA~ and $\gamma=\frac{4}{10}$.} 
\label{fig3}
\end{figure}

We next compare the EDL structures observed in the simulations of Coulomb lattice gas
with the ones found
for continuum ionic liquid, see Fig.~\ref{fig3}. We observe that, even at very dilute concentrations and weak Coulomb couplings, the lattice model does not agree completely with its continuum counterpart -- see Fig.~\ref{fig3}(a). 
This is somewhat surprising if we note that we are in the regime corresponding to aqueous electrolytes. This suggests that
lattice and continuum models will agree only for infinite dilution, $\gamma \to 0$,  and low surface charges.

\begin{figure}[h]
\begin{center}
\includegraphics[scale=0.3]{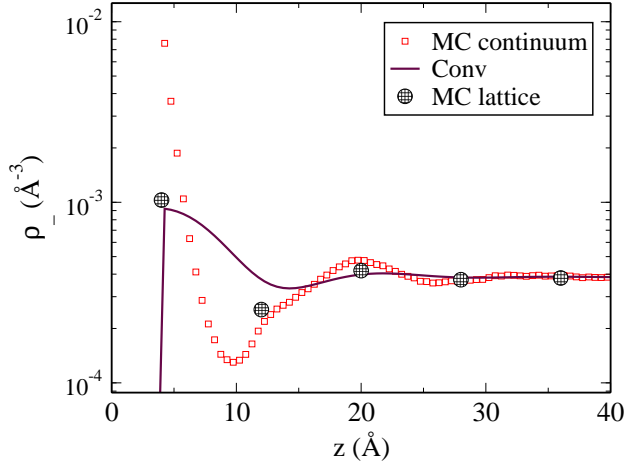}
\end{center}
\caption{Density profiles of anions near a positively charged EI. Square symbols are MC simulations of a continuum ionic fluid, while circles are the results of simulations of the Coulomb lattice gas. The line corresponds to the convolution of continuum MC profile with a Gaussian, see Eq.~(\ref{cov}). The surface charge on the electrode is $\sigma_s=0.10$C/m$^2$ . The parameters were chosen to correspond to typical RTILs, $\lambda_B=38.4$\AA~ and $\gamma=\frac{4}{10}$. }
\label{fig4}
\end{figure}

\begin{figure}[h]
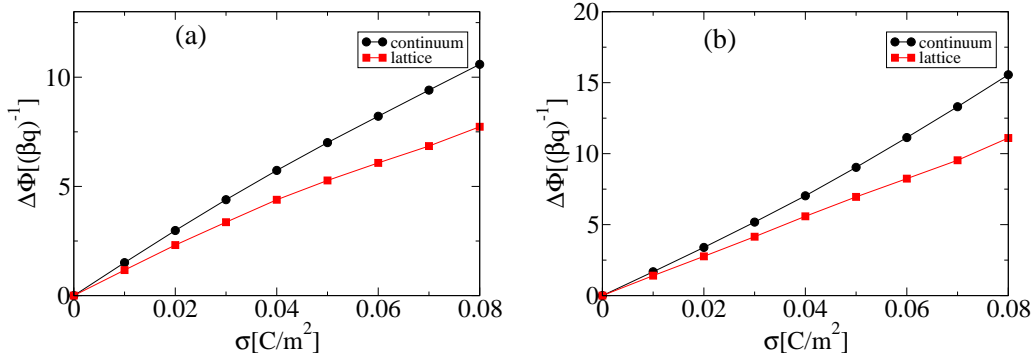

\begin{center}
\includegraphics[scale=0.25]{fig5a.eps}\hspace{0.3cm}
\includegraphics[scale=0.25]{fig5b.eps}
\end{center}
\caption{Potential difference as a function of superficial charge density. Circle correspond to the results of MC simulations of continuum model and squares of the  lattice Coulomb gas. The parameters are chosen to correspond to typical electrolytes in (a), $\lambda_B=7.2$\AA~ and $\gamma=\frac{1}{20}$, and typical RTILs in (b), $\lambda_B=38.4$\AA~ and $\gamma=\frac{4}{10}$.} 
\label{fig5}
\end{figure}

The discrepancy between lattice and continuum becomes more pronounced when we reach parameters characteristic or more realistic ionic liquids, see Fig.~\ref{fig3}(b). We see that, even though the lattice gas model captures qualitatively the structure of the continuum EDL, it fails quantitatively.  In particular, the contact densities are very different in the lattice and continuum models.  
At moderately high surface charge $\sigma_s$, see Fig.\ref{fig4}, the lattice model has  contact density an order of magnitude lower than the continuum system. These differences are due to the coarse grained structure of the lattice model, which effectively averages the microscopic details of the continuum fluid  over a lattice spacing. In order to test this hypothesis, we convolute the continuum ionic profile with a normalized Gaussian distribution with standard deviation equal to $R$:
\begin{equation}
\tilde{\rho} (z) = C \int_0^d dz' e^{-\frac{(z-z')^2}{2R^2}} \rho(z') \ ,
\label{cov}
\end{equation}
where $C$ normalizes the Gaussian over the range $d$, $\rho(z)$ is the ionic profile obtained from the continuum simulations, and $R$ is the lattice spacing. Although the choice of the convolution function is not unique, it will not makes a significant difference to the final result. Within the density functional theory, it is known that the appropriate coarse-graining has to be performed on the scale of one particle diameter, for the lattice model this translates to a lattice spacing. Furthermore, since the range of the Gaussian distribution is unlimited, it makes it easier to work out the convolution (and de-convolution) relations in the Fourier space. The results from the convolution relation (\ref{cov}) are shown in Figs.~\ref{fig3} and \ref{fig4}. We see that the convolution provides a fairly accurate map between ionic distributions found in the two models. The  agreement appears to be even better in the  RTILs regime.  This opens a possibility of extracting continuum density profiles from lattice model simulations, which are much simpler to perform then their continuum counterparts, using a deconvolution. As the ionic packing fraction becomes very large, the acceptance rate for single move displacements becomes very small, rendering the continuum MC simulations increasingly time demanding. The connection, Eq.~(\ref{cov}), between lattice and continuum models may allow one to estimate the real density profiles from the lattice simulations and vice-versa. Nevertheless, the process of deconvolution may be a complex one. This not withstanding, Figs.~\ref{fig3} and \ref{fig4} clearly show that the lattice model corresponds to a coarse-graining of the continuum system. This suggests that even though the lattice model can not account for the fine-grained details of a continuum system, it might be sufficient to study the more coare-grained thermodynamical properties of ionic liquids. To test this conjecture we calculated the potential difference between the surfaces as a function of the superficial charge density for the continuum and lattice models. Figs.\ref{fig5} shows that although there is a qualitative similarity between the two systems,  lattice model consistently underestimates the value of the potential difference.

\section{Conclusions}

We have studied density distributions of charged particles confined by oppositely charged flat surfaces. The parameter range under investigation was set to vary from aqueous electrolyte solutions to RTILs. This allowed us to explore the validity of the mPB theory in different regimes. mPB has been widely used to study ionic distributions and differential capacitance of ionic liquids. For system parameters typical of classic aqueous electrolytes (low ionic coupling and low ionic dilutions), mPB describes reasonably well the results of Coulomb lattice gas simulations.  For such parameters, however, the standard (not modified) PB equation performs equally well. As the ionic packing fraction becomes large -- regime characteristic of ionic liquids -- the structural features resulting from strong steric repulsion between ions appear, leading to the breakdown of mPB equation. Although such failure could have been anticipated  from the traditional PB approach (which completely neglects ionic size effects), the degree of accuracy of mPB  as applied to highly concentrated systems was not very clear. 
This is because finite size effects are partially taken into account in this equation 
through the excluded volume contribution to the underlying free energy. The steric effects, however, are included at the local density approximation (LDA) level, which is known to be inappropriate for dense systems.  The comparison of mPB equation with the simulations of a lattice Coulomb gas model on which this equation is based, clearly show the limitations of mPB theory.  The absence of electrostatic correlations and inclusion of the steric interactions between the ions only at 
the LDA level makes mPB equation inappropriate for studying RTILs. 

We have also compared the results of MC simulations of continuum and lattice versions of ionic liquid.  Although
qualitatively the results are similar, there are important quantitative difference in the density profiles 
of the two models.  In particular, the contact densities in the two models are very different.  In spite of this
quantitative difference, we find that the two density distributions can be connected semiquantitatively through a convolution relation. This suggests that in the lattice gas, the charge distributions of the continuum model are ``coarse grained" on the scale of lattice spacing.  This raises an important question if
lattice models can also be used to quantitative calculate the thermodynamic properties of continuum EDLs, and in particular their differential capacitance.  This will be the subject of the future work.

\section{Acknowledgments}
This work was partially supported by the CNPq, INCT-FCx, and by the US-AFOSR under the grant 
FA9550-16-1-0280.

\bibliography{ref}

\end{document}